\documentclass[pra,aps,twocolumn,showpacs,preprintnumbers,showkeys,superscriptaddress,english]{revtex4}

\usepackage{amssymb}
\usepackage{amsmath}
\usepackage{amscd}
\usepackage{babel}
\usepackage{latexsym}
\usepackage{epsfig}
\usepackage{bbm}

\newtheorem{corollary}{Corollary}
\newtheorem{lemma}{Lemma}
\newtheorem{theorem}{Theorem}
\newtheorem{proposition}{Proposition}

\newcommand{\qed}{{\hfill$\Box$}}
\newenvironment{proof}{\noindent \textbf{{Proof~} }}{\qed}
\def\bi{\begin{itemize}}
\def\ei{\end{itemize}}
\def\be{\begin{equation}}
\def\ee{\end{equation}}
\def\bea{\begin{eqnarray}}
\def\eea{\end{eqnarray}}
\def\ben{\begin{eqnarray*}}
\def\een{\end{eqnarray*}}

\def\>{\rangle}
\def\<{\langle}

\def\ra{\rightarrow}

\def\bK{{\bf K}}

\def\bS{{\bf S}}

\def\bbE{\mathbb{E}}

\newcommand{\ket}[1]{| #1 \rangle}

\newcommand{\proj}[1]{| #1 \>\!\< #1 |}

\DeclareMathOperator{\tr}{Tr}

\def\*{\star}

\def\tilde{\widetilde}
\def\bar{\overline}

        \def\cB{{\cal B}}
\def\cC{{\cal C}}

    \def\cH{{\cal H}}

        \def\cN{{\cal N}}

\def\cS{{\cal S}}        \def\cT{{\cal T}}

        \def\cW{{\cal W}}
\def\cX{{\cal X}}

\begin{document}

\title{Secret Keys Assisted Private Classical Communication Capacity over Quantum Channels}%
\author{Min-Hsiu~Hsieh}
\affiliation{Ming Hsieh Department of Electrical Engineering, University of Southern California, Los Angeles, CA 90089}%
\author{Zhicheng Luo}
\affiliation{Physics Department, University of Southern California, Los Angeles, CA 90089.}%
\author{Todd Brun}
\affiliation{Ming Hsieh Electrical Engineering Department, University of Southern California, Los Angeles, CA 90089}%

\begin{abstract}
We prove a regularized formula for the secret key-assisted capacity
region of a quantum channel for transmitting private classical
information. This result parallels the work of Devetak on
entanglement assisted quantum communication capacity \cite{DHW05RI}.
This formula provides a new family protocol, the private father
protocol, under the resource inequality framework that includes
private classical communication \it{without} secret key assistance
as a child protocol.
\end{abstract}


\maketitle

\section{Introduction}
\label{secI}%
Secret keys, by definition, refer to common randomness available to
a sender and receiver at distant locations while any other party has
absolutely no information about it. Generating secret keys requires
preserving secrecy from a third party \cite{AC93CR}. An
information-theoretic model in the classical setting is the
``wiretap channel" \cite{Wyner75Bell}, where the sender wants to
communicate with one legitimate receiver while keeping the
eavesdropper completely ignorant of the message sent. Private
communication can be achieved via encryption once secret keys are
generated. Secret keys are a valuable resource that can be used to
achieve information transmission tasks.

The above scenario has a quantum analogue, where secret keys are
generated over a quantum channel. The secret key generating protocol
has been proposed by several authors \cite{Lloyd96,SN96,DW03c}, and
in \cite{devetak03}, it has been shown that the capacity of a
quantum channel for transmitting private classical information is
the same as the capacity of the same channel for generating secret
keys. Furthermore, neither capacity is enhanced by public classical
communication in the forward direction. This raises the interesting
question of how these different resources interconvert in a quantum
information protocol, and was partially answered in
\cite{DHW03,DHW05RI,ADHW06FQSW}.

The formal treatment of quantitative interconversions between
nonlocal information processing resources is studied in
\cite{DHW05RI}, wherein such an asymptotically faithful conversion
is expressed as a \emph{resource inequality} (RI). These resource
inequalities are extremely powerful, and sometimes lead to new
quantum protocols \cite{DHW03}. For example, they allow us to relate
the \emph{family protocols} to several well-known quantum protocols
by direct application of teleportation or superdense coding, etc.

In this paper, we study the private classical communication capacity
over a quantum channel assisted by a secret key. We show that secret
keys are a useful nonlocal resource that can increase the private
classical communication capacity over quantum channels; however,
unlimited secret keys do not help. The trade-off between the rate of
secret key consumption and the rate of increased private classical
communication is presented quantitatively. Under the RI framework,
our protocol can be understood as a ``private father protocol'' due
to its similarity to the original father protocol. Furthermore, the
unassisted private classical communication capacity \cite{devetak03}
can be seen as a child protocol.

This paper is organized as follows. Section \ref{secII} contains
definitions, notation, and relevant background material. Section
\ref{secIII} contains statements and proofs of our main result. In
section \ref{secIV}, we rewrite our result under the RI framework,
and show how to recover the unassisted private classical capacity
from ours. We conclude in section \ref{secV}.

\section{Notation}
\label{secII}%
Consider a classical-quantum system $XQ$ in the state described by
an ensemble $\{p(x),\rho_x\}$ with $p(x)$ defined on $\cX$ and
$\rho_x$ being density operators on the Hilbert space $\cH_Q$ of
system $Q$. Such a state $\rho^{XQ}$ of systems $XQ$ can be
represented by the ``enlarged Hilbert space'' (EHS) representation:
$$\rho^{XQ}=\sum_{x}p(x)\proj{x}^{X}\otimes\rho^{Q}_{x},$$
where $X$ is a dummy quantum system and $\{\ket{x}:x\in\cX\}$ is an
orthonormal basis for the Hilbert space $\cH_X$ of system $X$. The
reduced density operators of systems $X$ and $Q$ are $\rho^X=\tr_Q
\rho^{XQ}=\sum_{x}p(x) \proj{x}$, and $\rho^Q=\tr_X\rho^{XQ}$
respectively. The von Neumann entropy of the quantum state $\rho^Q$
is $H(Q)_{\rho}=-\tr(\rho^Q\log\rho^Q)$. (We will omit the subscript
$\rho$ when the state is clear from the context.) Notice that the
von Neumann entropy of the dummy quantum system $X$ is equal to the
Shannon entropy of random variable $X$ whose probability
distribution is $p(x)$. The conditional entropy is defined as
\begin{equation}
\label{Hcond} H(Q|X)=H(QX)-H(X).
\end{equation}
It should be noted that conditioning on classical variables
(systems) amounts to averaging; therefore (\ref{Hcond}) is also
equal to
\begin{equation}
H(Q|X)=\sum_{x}p(x)H(Q)_{\rho_x}.
\end{equation}
The mutual information is $$I(X;Q)=H(X)+H(Q)-H(QX).$$

Next, we will briefly introduce definitions and properties of
typical sequences and subspaces \cite{NC00}.

Let $\cT_{X,\delta}^n$ denote the set of {\it typical sequences}
associated with some random variable $X$ such that for the
probability distribution $p$ defined on the set $\cX$
\begin{equation*}
\cT_{X,\delta}^n=\left\{x^n:\forall x \in\cX,
\left|\frac{N(x|x^n)}{n}-p(x)\right|\leq \delta\right\},
\end{equation*}
where $N(x|x^n)$ is the number of occurrences of $x$ in the sequence
$x^n:=x_1\cdots x_n$ of length $n$.

%Let $\cT_{X|U,\delta}^n(u^n)$ denote the set of {\it conditionally
%typical sequences} associated with some random variables $X$
%conditioned on $U$ such that for the conditional probability
%distribution $p(x|u)$ with $x\in\cX$ and $u\in\cU$
%\begin{align*}
%&\cT_{X|U,\delta}^n(u^n)=  \\ &\left\{x^n:\forall ux,
%\left|\frac{N((u,x)|(u^n,x^n))}{n}-\frac{N(u|u^n)}{n}p(x|u)\right|\leq\delta\right\}.
%\end{align*}

Assume the density operator $\rho^Q$ of system $Q$ has the following
spectral decomposition: $\rho^Q=\sum_y p(y)\proj{y}$. Then we can
define the {\it typical projector} as
\begin{equation}
\Pi_{Q,\delta}^n=\sum_{y^n\in\cT_{Y,\delta}^n}\proj{y^n},
\end{equation}%
where $\ket{y^n}$ is a state in $\cH_Q^{\otimes n}$. For a
collection of states $\{\rho_x$, $x\in\cX\}$, the {\it conditional
typical projector} is defined as
\begin{equation}
\Pi_{Q|X,\delta}^n(x^n)=\bigotimes_x \Pi_{Q_{|x},\delta}^{I_x},
\end{equation}%
where $I_x=\{i:x_i=x\}$ is the indicator and
$\Pi_{Q_{|x},\delta}^{I_x}$ denotes the tensor product of the
typical projector of the density operator $\rho_x^Q$ in the
positions given by the set $I_x$ with the identity everywhere else.

Fixing $\delta>0$, we will need the following properties of typical
subspaces and conditionally typical subspaces:
\begin{eqnarray}
\tr \sigma_{x^n}^Q \Pi^n_{Q|X,\delta}(x^n) & \geq & 1-\epsilon \\
\tr \sigma_{x^n}^Q \Pi^n_{Q,\delta(|\cX|+1)} & \geq & 1-\epsilon \\
\tr \Pi^n_{Q,\delta(|\cX|+1)} & \leq & \alpha \\
\Pi^n_{Q|X,\delta}(x^n) \sigma^Q_{x^n} \Pi^n_{Q|X,\delta}(x^n) &
\leq & \beta^{-1} \Pi^n_{Q|X,\delta}(x^n)
\end{eqnarray}
where $\alpha=2^{n[H(Q)+c\delta]}$ and $\beta=2^{n[H(Q|X)-c\delta]}$
for $\epsilon=2^{-nc'\delta^2}$ and some constants $c$ and $c'$.

Finally we need some facts about trace distances (taken from \cite{NC00}).
The trace distance between two density
operators $\rho$ and $\sigma$ can be defined as
$$\|\rho - \sigma\|_1 = \tr|\rho - \sigma|,$$
where $|A|\equiv\sqrt{A^{\dagger}A}$ is the positive square root of
$A^{\dagger}A$. The \emph{monotonicity} property of trace distance
is
\begin{equation}%
\|\rho^{RB} - \sigma^{RB}\|_1 \geq \|\rho^{B} - \sigma^{B}\|_1~.
\label{monodis}
\end{equation}%

\section{Main result}
\label{secIII}%
\subsection{Classical-quantum channels}%
We begin by defining our private classical communication protocol
for a $\{c\to qq\}$ channel from sender Alice to receiver Bob and
eavesdropper Eve. The channel is defined by the map $\cW: x \to
\sigma_{x}^{BE}$, with $x\in\cX$ and the state $\sigma_{x}^{BE}$
defined on a bipartite quantum system $BE$; Bob has access to
subsystem $B$ and Eve has access to subsystem $E$. Alice's task is
to transmit, by some large number $n$ uses of the channel $\cW$, one
of $\{0,1\}^{nR}$ equiprobable messages to Bob so that he can
identify the message with high probability while at the same time
Eve receives almost no information about the message. In addition,
Alice and Bob are given some private strings (secret keys), picked
uniformly at random from the set $\{0,1\}^{n R_s}$, before the
protocol begins. The inputs to the channel $\cW^{\otimes n}$ are
classical sequences $x^n \in\cX^{n}$ with probability $p^n(x^n)$.
The outputs of $\cW^{\otimes n}$ are density operators
$\sigma^{BE}_{x^n} =
\sigma^{BE}_{x_1}\otimes\cdots\otimes\sigma^{BE}_{x_n}$ living on
some Hilbert space $\cH^{B^nE^n}$. % In addition, they can generate a
%rate of $R$ bits/copy of secret communication, so Alice may send an
%arbitrary private string $m$ from the set $\{0,1\}^{n R}$ to Bob.

An $(n,R,R_s,\epsilon)$ \emph{secret key-assisted private channel
code} consists of
\begin{itemize}
    \item An encryption map $f: \{0,1\}^{n R} \times \{0,1\}^{n R_s}
    \ra\{0,1\}^{nR}$, i.e. $f$ generates an index random variable $K$
    uniformly distributed in $\{0,1\}^{nR}$ based on the classical
    message embodied in the random variable $M$ and the shared secret
    key embodied in the random variable $S$. Furthermore,
    $f(m,s_1)\neq f(m,s_2)$ for $s_1\neq s_2$ and $f(m_1,s)\neq f(m_2,s)$
    for $m_1\neq m_2$.
    %If the value of the secret key is $s\in\{0,1\}^{n R_s}$, Alice encrypts her private message
    %$m\in\{0,1\}^{n R}$ as the index $k = f(m,s)$;

    \item An encoding map $E: \{0,1\}^{n R} \ra \cX^n$. Alice encodes the index $k$ as
    $E(k)$ and sends it through the channel $\cW^{\otimes n}$, generating the state
    \begin{align}
    \Upsilon^{A_sB_sBE}& = \frac{1}{2^{nR_s}}\sum_{s\in\{0,1\}^{n R_s}}\proj{s}^{A_s} \otimes \proj{s}^{B_s} \otimes \nonumber\\
    &\frac{1}{2^{nR}}\sum_{m\in\{0,1\}^{n R}}\sigma_{E(f(m,s))}^{BE}
    \label{statebf}
    \end{align}

    \item A decoding POVM $\{\Lambda_{k'}\}_{k'\in\{0,1\}^{n R}}$, where $\Lambda_{k'}$ is a
    positive operator acting on $B$ and taking on values $k'$. Bob need to infer the index $k$
    through the POVM;

    \item A decryption map $g: \{0,1\}^{n R} \times \{0,1\}^{n R_s} \ra \{0,1\}^{n R}$, where $g(f(m,s),s)=m$, $\forall s,m$. This
    allows Bob to recover Alice's message as $m'=g(k',s)$ based on $k'$ and $s$;
\end{itemize}
such that
\begin{equation}
\label{cond1}%
\|\tilde{\Upsilon}^{BE} - \tau^{B} \otimes \sigma^E\|_1 \leq
\epsilon,
\end{equation}
where $\tilde{\Upsilon}^{BE}$ is the state of the subsystem $BE$
after Bob's decoding operation, and
$$
\tau^{B}=\frac{1}{2^{nR}}\sum_m\proj{m}^B
$$
contains the private classical information that is decoupled from
Eve's state $\sigma^E$.

A rate pair $(R,R_s)$ is called \emph{achievable} if for any
$\epsilon,\delta>0$ and sufficiently large $n$ there exists an
$(n,R-\delta,R_s+\delta,\epsilon)$ private channel code. The private
capacity region $C_{PF}(\cW)$ is a two-dimensional region in the
$(R,R_s)$ plane with all possible achievable rate pairs $(R,R_s)$.

We now state our main theorem.
\begin{theorem}
\label{pf}%
The private channel capacity region $C_{PF}(\cW)$ is given by
\begin{equation}
\label{cp}%
C_{PF}(\cW)=\bar{\bigcup_{n=1}^{\infty}\frac{1}{n}\tilde{C}_{PF}^{(1)}(\cW^{\otimes
n})},
\end{equation}
where the notation $\bar{Z}$ means the closure of a set $Z$ and
$\tilde{C}^{(1)}_{PF}(\cW)$ is the set of all $R_s\geq 0$, $R\geq 0$
such that
\begin{eqnarray}
R & \leq & I(X;B)_\sigma-I(X;E)_\sigma+R_s  \label{thmcond1}\\
R & \leq & I(X;B)_\sigma, \label{thmcond2}
\end{eqnarray}
where $BE|X$ is given by $\cW$ and $\sigma$ is of the form
$$\sigma^{XBE}=\sum_{x}p(x)\proj{x}^{X}\otimes\sigma^{BE}_{x}.$$
\end{theorem}
Proving that the right hand side of (\ref{cp}) is achievable is
called the {\it direct coding theorem}, whereas showing that it is
an upper bound is called the {\it converse}.

For the direct coding part, we will need the following lemma from \cite{DHW06}, a quantum
generalization of the covering lemma in \cite{LD06CS}.
\begin{lemma}[Covering Lemma]
\label{covering} We are given an ensemble $\{p(x),\sigma_x\}_{x\in
\cX}$ with average density operator $\sigma=\sum_{x \in
\cX}p(x)\sigma_x$. Assume the existence of projectors $\Pi$ and
$(\Pi_{x})_{x\in\cX}$ with the following properties ($\forall x\in
\cX$):
\begin{eqnarray*}
\tr \sigma_x \Pi_x & \geq & 1-\epsilon, \label{cover1} \\
\tr \sigma_x \Pi & \geq & 1-\epsilon, \label{cover2}\\
\tr \Pi & \leq & \alpha, \label{cover3} \\
\Pi_x \sigma_x \Pi_x & \leq & \beta^{-1} \Pi_x. \label{cover4}
\end{eqnarray*}
In addition, we require $\Pi_x$ and $\sigma_x$ to commute for all
$x$. The {\it obfuscation error} of a set $\cS\subseteq \cX$ is
defined as
$$oe(\cS)=\left\|\frac{1}{|\cS|}\sum_{x\in\cS}\sigma_x-\sigma \right\|_1,$$
and is an upper bound on the probability of distinguishing the fake
average from the real one. Define the set $\cC=(X_s)_{s\in[\bS]}$,
where $X_s$ is a random variable chosen independently according to
the distribution $p$ on $\cX$, and $\bS=\lceil \gamma^{-1}
\alpha/\beta \rceil$ for some $0<\gamma<1$. Then
\begin{equation}
\Pr\{oe(\cC)\geq 2\epsilon+19\sqrt{\epsilon}\}\leq
2\alpha\exp(-\kappa_0\epsilon^3/\gamma).
\end{equation}
\end{lemma}

\medskip

\begin{corollary}
Consider an ensemble $\{p^n(x^n),\sigma_{x^n}^E\}_{x^n\in \cX^n}$
with average density operator
$\sigma^E=\sum_{x^n}p^n(x^n)\sigma_{x^n}^E$, let random variables
$X_1, X_2,...,X_\bS$ all be independently distributed according to
$p^n$ and $\cC=(X_s)_{s\in[\bS]}$. Then for all $\epsilon, \delta>0$
and sufficiently large $n$,
\begin{equation}
\Pr\{oe(\cC)\geq 2\epsilon+19\sqrt{\epsilon}\}\leq
2\alpha\exp(-\kappa_0\bS\epsilon^3\beta/\alpha).
\end{equation}
where $\alpha = 2^{n[H(E)_\sigma+c\delta]}$, $\beta =
2^{n[H(E|X)_\sigma-c\delta]}$, $\bS=2^{nI(X;E)_\sigma+3c\delta}$,
and
$$
oe(\cC) =
\left\|\frac{1}{\bS}\sum_{s\in[\bS]}\sigma^E_{X_s}-\sigma^E
\right\|_1~.
$$
\label{cor:covering}
\end{corollary}

\begin{proof}
We can relate to Lemma $\ref{covering}$ through the identifications: $\cX \ra \cX^n$,
$\sigma_x \ra \sigma_{x^n}$, $p \ra p^n$, $\sigma \ra \sigma^E$,
$\Pi \ra \Pi^n_{E, \delta(|\cX|+1)}$ and $\Pi_{x} \ra \hat{\Pi}^n_{E|X, \delta}(x^n)$ with
$$
\hat{\Pi}^n_{E|X, \delta}(x^n)= \left\{\begin{array}{cc}
\Pi^n_{E|X, \delta}(x^n) \,\,  & x^n \in \cT^n_{X, \delta} \\
0 \,\, & {\rm otherwise.}
\end{array} \right.
$$

The four conditions now read (for all $x^n\in\cX^n$),
\begin{align}
\tr \sigma_{x^n}^E \hat{\Pi}^n_{E|X,\delta}(x^n) & \geq  1-\epsilon , \label{cor:c1}\\
\tr \sigma_{x^n}^E \Pi^n_{E,\delta(|\cX|+1)} & \geq  1-\epsilon, \label{cor:c2}\\
\tr \Pi^n_{E,\delta(|\cX|+1)} & \leq  \alpha, \label{cor:c3}\\
\hat{\Pi}^n_{E|X, \delta}(x^n) \sigma_{x^n}^E \hat{\Pi}^n_{E|X, \delta}(x^n) &
\leq  \beta^{-1} \hat{\Pi}^n_{E|X, \delta}(x^n). \label{cor:c4}
\end{align}
These follow from the properties of typical subspaces and conditionally typical subspaces
mentioned before.
\end{proof}
\medskip

We will also need the Holevo-Schumacher-Westmoreland (HSW) theorem
\cite{Hol98, SW97}.

\begin{proposition}[HSW theorem]
Given an ensemble
$$
\sigma^{XB} = \sum_{x \in \cX} p(x) \proj{x}^X \otimes \sigma_x^B,
$$
and integer $n$, consider the encoding map $E:[\bK] \rightarrow
\cX^n$ given by $E(k) = X_k$, where $k\in[\bK]:1,2,\cdots,\bK$, and
$\{X_k\}$ are random variables chosen according to the i.i.d.
distribution $p^n$. For any $\epsilon, \delta>0$ and sufficiently
large $n$, there exists a decoding POVM set $(\Lambda_k)_{k \in
[\bK]}$ on $B$ for the encoding map $E$ with $\bK =
2^{n[I(X;B)-2(c+c'\delta)\delta]}$, and some $c$, such that for all
$k$,
$$
\bbE \sum_{k'} | \pi({k'|k})  - \delta(k,k') | \leq \epsilon~.
$$
Here $\pi({k'|k})$ is the probability of decoding $k'$ conditioned
on $k$ having been encoded:
\begin{equation}
 \pi({k'|k}) = \tr( \Lambda_{k'}  \sigma_{E(k)}),
\end{equation}
$\delta(s,s')$ is the delta function
and the expectation is taken over the random encoding.
\label{prop:HSW}
\end{proposition}

%\medskip

\begin{figure}
\centering
\includegraphics[scale=0.75]{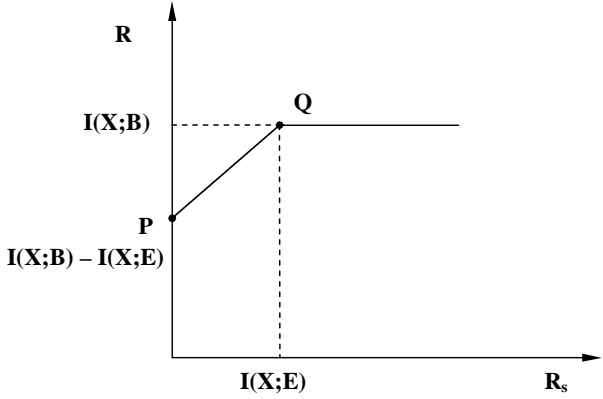}
\caption{Private classical communication capacity region of $\{c\to
qq\}$ channel when assisted by a pre-shared secret key.}
\label{fig:region}
\end{figure}

Now we are ready to prove the direct coding theorem.

\begin{proof}[{\bf direct coding theorem}]

The capacity region is shown in Fig.~\ref{fig:region}. This
trade-off region includes two limit points $P$ and $Q$. When $R_s=0$
(Point P), the private classical capacity of $\cW$ is equal to
$I(X;B)-I(X;E)$. This is the well-known private classical
communication capacity proved in \cite{devetak03}. In our case, it
suffices to prove Point Q is optimal; that is, the achievability of
the rate pair $(R,R_s) = (I(X;B), I(X;E))$. The idea of the proof is
as follows: instead of sacrificing $nI(X;E)$ bits of classical
message to randomize Eve's knowledge of the state, Alice and Bob use
a pre-shared secret key to do so. For all $\epsilon,\delta>0$ and
sufficiently large $n$, we show below that a private information
transmission rate of $I(X;B)$ is achievable if Alice and Bob consume
a pre-shared secret key with rate $I(X;E)$.

Fix $\epsilon, \delta > 0$ and a sufficiently large $n$. Consider
the ensemble $\{p^n(x^n),\sigma_{x^n}^{BE}\}$ of the channel output
$\cW^{\otimes n}$. There exists an encoding map $E:K\ra X_K$ for
Alice on the encryption output $K=f(M,S)$ where $X_K$ is i.i.d. with
distribution $p^n$, $M$ represents the classical message taken
values from $\{0,1\}^{nR}$, and $S$ represents the pre-shared secret
key values taken from $\{0,1\}^{nR_s}$. Here $\{X_{K}\}$ serves as a
HSW code.

In the following, we will explicitly use $f(m,s)$ instead of its
index $k$. For each $m\in\{0,1\}^{nR}$, define $\cC_m =
(X_{f(m,s)})_{s\in[2^{nR_s}]}$. $\cC_m$ works as a covering code %We
%will conclude the proof by ``derandomizing'' the code, i.e. showing
%that a particular realization of the random $X_{f(m,s)}$ exists with
%suitable properties.
%
as define in Corollary $\ref{cor:covering}$. Choose $R_s =
I(X;E)+3(c+c'\delta)\delta$. For any $m\in\{0,1\}^{nR}$, define the
logic statement $\ell_m$ by $oe(\cC_m)\leq
2\epsilon+19\sqrt{\epsilon}$, where
$$
oe(\cC_m) = \left\|\frac{1}{2^{nR_s}}\sum_{s}\sigma^E_{X_{f(m,s)}}-\sigma^E \right\|_1~,
$$
$\sigma^E = \sum_{x^n}p^n(x^n)\sigma^E_{x^n}$ and
$\sigma^E_{x^n}=\tr_B \sigma^{BE}_{x^n}$. By Corollary
$\ref{cor:covering}$,
\begin{equation}
\label{pr}%
\Pr\{{\rm not}\,\, \ell_m\} \leq
2\alpha\exp(-\kappa_02^{nR_s}\epsilon^3\beta/\alpha)~,\ \ \forall m.
\end{equation}
The probability of (\ref{pr}) can be made $\leq \epsilon 2^{-nR}$
when $n$ is sufficient large.

We now invoke Proposition $\ref{prop:HSW}$. Choose $R =
I(X;B)-2(c+c'\delta)\delta$. There exists a POVM $(\Lambda_{k'})_{k'
\in \{0,1\}^{nR}}$ acting on $B$ such that for all $k$,
\begin{equation}
\bbE \sum_{k'} | \pi({k'|k})  - \delta(k,k') | \leq \epsilon~.
\label{POVM}
\end{equation}
%Here $\pi({k'|k})$ describes the noise experienced in conveying $k$ to Bob.
After Bob performs the POVM, the state ($\ref{statebf}$) becomes
\begin{align*}
\hat{\Upsilon} = \frac{1}{2^{nR_s}}&\sum_{s}\proj{s}^{A_s} \otimes \proj{s}^{B_s} \otimes
\frac{1}{2^{nR}}\sum_{m,k'}\pi(k'|f(m,s))\\
&\proj{k'}^B\otimes \sigma_{X_{f(m,s)}}^{E}~,
\end{align*}
which is close to
\begin{align*}
\hat{\Upsilon}_0 = &\frac{1}{2^{nR_s}}\sum_{s}\proj{s}^{A_s} \otimes \proj{s}^{B_s}  \\
&\otimes \frac{1}{2^{nR}}\sum_{m}\proj{f(m,s)}^B\otimes
\sigma_{X_{f(m,s)}}^{E}
\end{align*}
in the sense that $\bbE\|\hat{\Upsilon} - \hat{\Upsilon}_0\|_1\leq
\epsilon$ by condition ($\ref{POVM}$).

Bob applies the decryption map $g$ to his system $B$, resulting in a
state $\tilde{\Upsilon}^{A_sB_sBE}$. By the monotonicity of trace
distance ($\ref{monodis}$), we have
$$%
\bbE\|\tilde{\Upsilon}^{BE} - \tilde{\Upsilon}_0^{BE}\|_1\leq
\epsilon~,
$$
where
$$
\tilde{\Upsilon}_0^{BE} = \frac{1}{2^{nR}}\sum_m \proj{m}^B\otimes
\frac{1}{2^{nR_s}} \sum_s\sigma_{X_{f(m,s)}}^{E}~.
$$
By the Markov inequality, $\Pr\{{\rm not}\,\,\ell_0\}\leq
\sqrt{\epsilon}$, where $\ell_0$ is the logic statement
\begin{equation}
\label{req}%
\|\tilde{\Upsilon}^{BE} - \tilde{\Upsilon}_0^{BE}\|_1\leq
\sqrt{\epsilon}.
\end{equation}
By the union bound,
$$
\Pr\{\text{not}\
(\ell_0\&\ell_1\&\cdots\&\ell_{|m|})\}\leq\sum_{i=0}^{2^{nR}}\Pr\{\text{not}\
\ell_i\} \leq \epsilon+\sqrt{\epsilon}.
$$
Hence there exists a specific choice of $\{X_{f(m,s)}\}$, say
$\{x_{f(m,s)}\}$, for which all these conditions are satisfied.
Consequently,
\begin{align*}
\|\tilde{\Upsilon}^{BE} - \tau^B\otimes \sigma^E\|_1 &\leq
\|\tilde{\Upsilon}^{BE} - \tilde{\Upsilon}_0^{BE}\|_1 \\
&+ \|\tilde{\Upsilon}_0^{BE} - \tau^B\otimes \sigma^E\|_1 \\
&\leq 2\epsilon + 20\sqrt{\epsilon}~.
\end{align*}
as claimed.
\end{proof}

\begin{proof}[{\bf converse}]%

We shall prove that, for any $\delta,\epsilon >0$ and sufficiently
large $n$, if an $(n,R,R_s,\epsilon)$ secret keys assisted private
channel code has rate $R$ then (\ref{thmcond1}) and (\ref{thmcond2})
hold.

\begin{figure}
\centering
\includegraphics[scale=0.45]{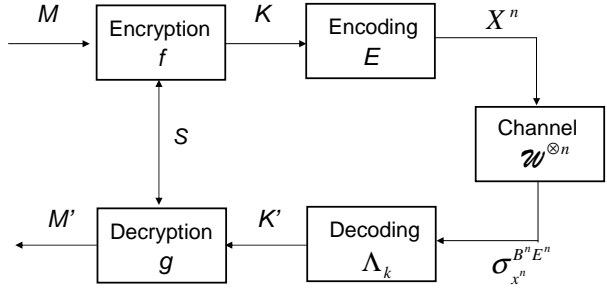}
\caption{Private classical communication protocol assisted by
pre-shared secret key.} \label{fig:converse}
\end{figure}
The private classical communication protocol is shown in
Fig.~\ref{fig:converse}
\begin{eqnarray*}
nR & = & H(K) \\
&= & I(K;K')+H(K|K') \\
&\leq& I(K;K')+1+n\epsilon \log |\cX|,
\end{eqnarray*}
where the last inequality follows from Fano's inequality:
$$H(K|K')\leq 1+\Pr\{K\neq K'\}nR,$$
and $\Pr\{K\neq K'\}\leq \epsilon$ is guaranteed by the HSW theorem.
Hence,
\begin{eqnarray}
I(K;K') & \leq & I(K;B^n) \\
& \leq & I(X^n;B^n),
%& \leq & I(K;B^n|S)-I(T;E^n|S) + R_s +\epsilon,
\end{eqnarray}
where the first inequality follow from the data processing
inequality while the second inequality comes from the Markov
condition $K\ra X^n\ra B^nE^n$. We then have
\begin{equation}
R-\delta \leq \frac{1}{n} I(X^n;B^n),
\end{equation}
where without loss of generality $\epsilon\leq
\frac{\delta}{6\log|\cX|}$ and $n\geq \frac{2}{\delta}$. This proves
(\ref{thmcond2}).

On the other hand,
\begin{align}
I(MS;M'S) & = I(MS;M'|S)+I(S;MS) \nonumber  \\
&\leq I(K;M'|S)+H(S) \label{coninv1}\\
&\leq I(K;B^n|S)+H(S) \label{coninv2}\\
& \leq I(X^n;B^n|S)+H(S) \label{coninv3}
\end{align}
where (\ref{coninv1}) follows from $I(MS;M'|S)=I(K;M'|S)$ and
$I(S;MS)\leq H(S)$, (\ref{coninv2}) follows from data processing
inequality, and (\ref{coninv3}) follows from the Markov condition
$K\to X^n\to B^nE^n$. Furthermore, (\ref{req}) guarantees that
\begin{align}
\epsilon & \geq I(M;E^n|S)\\
& =  I(K;E^n|S) \\
& \geq  I(X^n;E^n|S). \label{con2}
\end{align}
Combining (\ref{coninv3}) and (\ref{con2}) gives
\begin{align}
I(M;M') & \leq I(MS;M'S) \\
& \leq I(X^n;B^n|S) - I(X^n;E^n|S) \\
&+H(S) +\epsilon. \label{coninv4}
\end{align}
Hence
\begin{align}
nR &= H(M) \\
& =I(M;M')+H(M|M') \\
& \leq I(M;M') +1 +n\epsilon \log |\cX|, \label{coninv23}
\end{align}
where (\ref{coninv23}) follows from the Fano's inequality. Choosing
$\epsilon\leq \frac{\delta}{6\log|\cX|}$ and $n\geq
\frac{2}{\delta}$, (\ref{coninv4}) and (\ref{coninv23}) give
\begin{align}
R-\delta & \leq \frac{1}{n}\left[ I(X^n;B^n|S) - I(X^n;E^n|S) +
H(S)\right]  \\
& = \frac{1}{n}\left[ I(X^n;B^n|S) - I(X^n;E^n|S)\right] +R_s,
\end{align}
where $R_s= \frac{H(S)}{n}$. Since we can write
\begin{equation}
\frac{1}{n}\left[ I(X^n;B^n|S) - I(X^n;E^n|S)\right],
\end{equation}
as the average with respect to the distribution of $S$, and $K_s\to
X_s^n\to B^n_sX^n_s$ holds for each $s$, we can choose a particular
value of $s$ that maximizes (\ref{thmcond1}).

\end{proof}

\subsection{Generic quantum channels}
Suppose now that Alice and Bob are connected by a noisy quantum
channel $\cN:\cB(\cH_A') \to \cB(\cH_{B})$, where $\cB(\cH)$ denotes
the space of bounded linear operators on $\cH$. Let
$U_{\cN}:\cB(\cH_A') \to \cB(\cH_{BE})$ be an isometry extension of
$\cN$ that includes the unobserved environment $E$ which is
completely under the control of the eavesdropper Eve. Theorem
\ref{pf} then can be rewritten as the following

\begin{theorem}
\label{pf_2}%
The private channel capacity region $C_{PF}(\cN)$ is given by
\begin{equation}
\label{cp2}%
C_{PF}(\cN)=\bar{\bigcup_{n=1}^{\infty}\frac{1}{n}\tilde{C}_{PF}^{(1)}(\cN^{\otimes
n})},
\end{equation}
where the notation $\bar{Z}$ means the closure of a set $Z$ and
$\tilde{C}^{(1)}_{PF}(\cN)$ is the set of all $R_s\geq 0$, $R\geq 0$
such that
\begin{eqnarray}
R & \leq & I(A;B)_\sigma-I(A;E)_\sigma+R_s  \label{thm2cond1}\\
R & \leq & I(A;B)_\sigma, \label{thm2cond2}
\end{eqnarray}
where $\sigma$ is of the form
$$\sigma^{ABE}= U_{\cN}^{A'\to BE}(\ket{\psi}^{AA'}),$$
for some pure input state $\ket{\psi}^{AA'}$ whose reduced density
operator $\rho^{A'}=\sum_{x}p(x)\rho_x$ and $U_{\cN}:A'\ra BE$ is an
isometric extension of $\cN$.
\end{theorem}

With the spectral decomposition of the input state
$\rho^{A'}=\sum_{x}p(x)\rho_x$, each $U_\cN$ induces a corresponding
$\{c\to qq\}$ channel. Therefore, the results of the previous
section can be directly applied here.

\section{Private Father Protocol}
\label{secIV}%
In this section, we will phrase our result using the theory of
resource inequalities developed in \cite{DHW03}. The channel $\cN:A
\to B$ assisted by some rate $R_s$ of secret key shared between
Alice and Bob was used to enable a rate $R$ of secret communication
between Alice and Bob. This is written as
\begin{equation}
\label{RI1} \langle\cN\rangle+R_s[cc]^*\geq R[c\to c]^*.
\end{equation}
This resource inequality holds iff $(R_s,R)\in C_{PF}(\cN)$, with
$C_{PF}(\cN)$ given in Theorem \ref{pf_2}. The ``if'' direction,
i.e. the direct coding theorem, followed from the ``corner points''
\begin{equation}
\label{RI} \langle\cN\rangle+I(A;E)[cc]^*\geq I(A;B)[c\to c]^*.
\end{equation}
This resource inequality (\ref{RI}) is called the \emph{private
father protocol} due to its similarity of the father protocol in
\cite{DHW03}.

We can recover the unassisted private channel capacity result in
\cite{devetak03}:
\begin{equation}
\label{RI2} \langle\cN\rangle\geq I_c(A\rangle B)[c\to c]^*.
\end{equation}
This resource inequality can be obtained by appending the following
noiseless resource inequality
\begin{equation}
[c\to c]^* \geq[cc]^*
\end{equation}
to the output of (\ref{RI}).

\section{Conclusion}%
\label{secV}%

In this paper, we have found a regularized expression for the secret
keys assisted capacity region $C_{PF}(\cN)$ of a quantum channel
$\cN$ for transmitting private classical information. Our result
shows that secret key are a valuable nonlocal resource for
transmitting private information. One interesting problem is to
investigate how secret keys can be applied in other quantum
protocols. For example, it might be plausible that the entanglement
generation protocol could be boosted by secret keys. However, the
result seems unlikely. In particular, it is impossible to construct
a secret key-assisted entanglement generation protocol by simply
coherifying the protocol proposed in this paper. Another open
problem is to obtain a single-letterized formula of Theorem
\ref{pf}.

\section*{Acknowledgment}
We are grateful to Igor Devetak for valuable discussions. MH and ZL
were supported by NSF grant no.~0524811. TAB was supposed by NSF
grant no.~0448658.

\bibliography{../Ref}
\bibliographystyle{unsrt}
\end{document}